\documentclass[prc,twocolumn,nofootinbib,superscriptaddress,showpacs,aps]{revtex4-1}
\usepackage{amsmath}
\usepackage{amssymb}
\usepackage{mathtools}
\usepackage{graphicx}
\usepackage{dcolumn,dsfont}
\usepackage{bm,braket}
\usepackage{hyperref}
\usepackage{isotope}
\usepackage{xspace}
\usepackage{color}
\usepackage{multirow}
\usepackage{tabularx}
\usepackage{subfigure} 
\usepackage{blkarray}
\usepackage{relsize}
\usepackage{bbm}
\usepackage{pbox}
\usepackage{booktabs}
\usepackage{makecell}
\usepackage{ulem}
\usepackage[activate={true,nocompatibility},final,tracking=true,
kerning=true,spacing=true,factor=1100,stretch=10,shrink=10]{microtype}

\usepackage[all]{nowidow}
\hypersetup{
     colorlinks   = true,
     linkcolor = cyan,
     citecolor = cyan,
     menucolor = black,
     urlcolor = cyan  
}

\begin{document}
\renewcommand{\vec}[1]{\mathbf{#1}}
\newcommand{\clebschG}[6]{\mathcal{C}_{#1 #2 #3 #4}^{#5 #6}}  
\newcommand{\kF}{k_{\rm F}}
\newcommand{\MeV}{\,\mathrm{MeV}}
\newcommand{\GeVi}{\,\mathrm{GeV}^{-1}}
\newcommand{\fm}{\,\mathrm{fm}}
\newcommand{\fmi}{\,\mathrm{fm}^{-1}}
\newcommand{\fmiq}{\,\mathrm{fm}^{-3}}
\newcommand{\clebschGA}[6]{
  \left(
  \begin{array}{cc|c}
  #1 & #3 & #5 \\
  #2 & #4 & #6
  \end{array}
  \right)
}
\newcommand{\sixJsymbsixA}[6]{
	\begin{Bmatrix}
  #1 & #3 & #5 \\
  #2 & #4 & #6
	\end{Bmatrix}
}  
\newcommand{\nineJsymbA}[9]{
	\begin{Bmatrix}
  #1 & #4 & #7 \\
  #2 & #5 & #8\\
  #3 & #6 & #9 
	\end{Bmatrix}
}  

\newcommand{\oneSzero}{$\prescript{1\!}{}{S}_0$\xspace}
\newcommand{\onePone}{$\prescript{1\!}{}{P}_1$\xspace}
\newcommand{\threeSone}{$\prescript{3\!}{}{S}_1$\xspace}
\newcommand{\threeDone}{$\prescript{3\!}{}{D}_1$\xspace}
\newcommand{\threeSDone}{$\prescript{3\!}{}{S}_1\!-\!\prescript{3\!}{}{D}_1$\xspace}
\newcommand{\threePzero}{$\prescript{3\!}{}{P}_0$\xspace}
\newcommand{\threePone}{$\prescript{3\!}{}{P}_1$\xspace}
\newcommand{\threePtwo}{$\prescript{3\!}{}{P}_2$\xspace}
\newcommand{\threePFtwo}{$\prescript{3\!}{}{P}_2\!-\!\prescript{3\!}{}{F}_2$\xspace}
\newcommand{\threeFtwo}{$\prescript{3\!}{}{F}_2$\xspace}
\newcommand{\oneDtwo}{$\prescript{1\!}{}{D}_2$\xspace}
\newcommand{\oneGfour}{$\prescript{1\!}{}{G}_4$\xspace}

\newcommand{\ptilde}{\widetilde{p}} 
\newcommand{\ptildep}{\widetilde{p}'} 
\newcommand{\rtilde}{\widetilde{r}}
\newcommand{\WEA}{Weinberg eigenvalue analysis}

\renewcommand\theadalign{lt}
\renewcommand\theadgape{\Gape[4pt]}
\renewcommand\cellgape{\Gape[4pt]}

\title{Weinberg eigenvalues for chiral nucleon-nucleon interactions}

\author{J.\ Hoppe}
\email[Email:~]{jhoppe@theorie.ikp.physik.tu-darmstadt.de}
\affiliation{Institut f\"ur Kernphysik, Technische Universit\"at Darmstadt, 64289 Darmstadt, Germany}
\affiliation{ExtreMe Matter Institute EMMI, GSI Helmholtzzentrum f\"ur Schwerionenforschung GmbH, 64291 Darmstadt, Germany}

\author{C.\ Drischler}
\email[Email:~]{christian.drischler@physik.tu-darmstadt.de}
\affiliation{Institut f\"ur Kernphysik, Technische Universit\"at Darmstadt, 64289 Darmstadt, Germany}
\affiliation{ExtreMe Matter Institute EMMI, GSI Helmholtzzentrum f\"ur Schwerionenforschung GmbH, 64291 Darmstadt, Germany}

\author{R.\ J.\ Furnstahl}
\email[Email:~]{furnstahl.1@osu.edu}
\affiliation{Department of Physics, The Ohio State University, Columbus, OH 43210, USA}

\author{K.\ Hebeler}
\email[Email:~]{kai.hebeler@physik.tu-darmstadt.de}
\affiliation{Institut f\"ur Kernphysik, Technische Universit\"at Darmstadt, 64289 Darmstadt, Germany}
\affiliation{ExtreMe Matter Institute EMMI, GSI Helmholtzzentrum f\"ur Schwerionenforschung GmbH, 64291 Darmstadt, Germany}
 
\author{A.\ Schwenk}
\email[Email:~]{schwenk@physik.tu-darmstadt.de}
\affiliation{Institut f\"ur Kernphysik, Technische Universit\"at Darmstadt, 64289 Darmstadt, Germany}
\affiliation{ExtreMe Matter Institute EMMI, GSI Helmholtzzentrum f\"ur Schwerionenforschung GmbH, 64291 Darmstadt, Germany}
\affiliation{Max-Planck-Institut f\"ur Kernphysik, Saupfercheckweg 1, 69117 Heidelberg, Germany}

\begin{abstract}
We perform a comprehensive Weinberg eigenvalue analysis of a
representative set of modern nucleon-nucleon interactions derived
within chiral effective field theory. Our set contains local,
semilocal, and nonlocal potentials, developed by Gezerlis, Tews
\textit{et~al.}~(2013); Epelbaum, Krebs, and Mei{\ss}ner~(2015); and
Entem, Machleidt, and Nosyk~(2017) as well as Carlsson, Ekstr\"om
\textit{et~al.}~(2016), respectively. The attractive eigenvalues show
a very similar behavior for all investigated interactions, whereas the
magnitudes of the repulsive eigenvalues sensitively depend on the
details of the regularization scheme of the short- and long-range
parts of the interactions. We demonstrate that a direct comparison of
numerical cutoff values of different interactions is in general
misleading due to the different analytic form of regulators; for
example, a cutoff value of $R=0.8$~fm for the semilocal interactions
corresponds to about $R=1.2$~fm for the local interactions.  Our
detailed comparison of Weinberg eigenvalues provides various insights
into idiosyncrasies of chiral potentials for different orders and
partial waves. This shows that Weinberg eigenvalues could be used as a
helpful monitoring scheme when constructing new interactions.
\end{abstract}


\maketitle

\section{Introduction}
\label{sec:intro}

Chiral effective field theory (EFT) has become the standard method to
generate microscopic nuclear Hamiltonians for few- and many-body
calculations.  The dominant implementation is based on nucleon and
pion degrees of freedom (i.e., without explicit delta resonances) and
an organization dictated by the EFT power counting known as Weinberg
counting.  This specifies a diagrammatic expansion for inter-nucleon
potentials, which has been described in detail in several reviews
(e.g., see Refs.~\cite{Epel09RMP,Mach11PR}). But while the
diagrammatic content is prescribed, a potential requires specifying an
ultraviolet regularization scheme with an associated scale parameter
or possibly different parameters in separate many-body sectors.  Such
a scheme includes additional freedom in choosing the functional form
of the regulator function. Thus, there is an infinite variety of
candidate potentials to describe low-energy nuclear phenomena.
 
Up to a few years ago, a particular chiral EFT nucleon-nucleon (NN)
potential, specified over a decade ago in Ref.~\cite{Ente03EMN3LO} and
supplemented with the leading three-nucleon (3N) interaction, was used
in almost all many-body calculations (however, with different choices
for 3N regulators and fits). Improvements in many-body methods and the
advance of high-performance computing has enabled application to a
wide variety of nuclear systems (e.g., see
Refs.~\cite{Hage14rev,Hebe15ARNPS,Hage16NatPhys,Herg16PR}).  While
there have been notable phenomenological successes, the improved
precision and reach of these calculations have manifested deficiencies
in the Hamiltonian.  As a result, various groups have revisited the
construction and fitting of chiral potentials to better realize the
EFT advantages of systematic order-by-order improvement with
quantifiable errors.

Several different families (schemes) of nuclear interactions using
Weinberg counting have been introduced, with a variety of parameter
estimation methods used to fit the low-energy constants to nuclear
data.  These can be classified according to the regulator
implementation (see Sec.~\ref{sec:NN_interactions}) as local,
semilocal, or nonlocal, with broad freedom to choose the functional
form of the regulator within each category.  The NN interaction has
been pushed to fifth order in Weinberg counting
(``next-to-next-to-next-to-next-to-leading order'' or
N$^4$LO)~\cite{Ente14NNn4lo,Epel15NNn4lo,Ente17EMn4lo}, although for
consistency with 3N interactions various other lower order NN
interactions are available and have been applied.  In principle, these
interactions should all be capable of describing the same phenomena,
but in practice the detailed differences can be important.  While
effects of the regulator (so-called regulator artifacts) at a
given order in the expansion are supposed to be removed systematically
at higher orders, actual calculations show significant influence of
artifacts on the EFT convergence pattern.  In this work, we apply the
eigenvalue analysis methods developed by Weinberg~\cite{Wein63quasip}
(see also Refs.~\cite{Jost51scatstat,Kohn54ConvBorn,Meet62scattIOp})
to compare several sets of chiral NN potentials.

The Weinberg eigenvalue analysis is a versatile diagnostic tool to
quantify the perturbativeness of nuclear interactions and provide insight
into the physics of individual partial-wave channels.  Originally,
Weinberg developed this method in the early 1960s while working to
understand bound states in nonrelativistic quantum mechanics (as a
warm-up to understanding composite particles in quantum field theory)
and how to introduce quasiparticles to cure nonconvergent Born
series~\cite{Wein62elpart,Wein63quasip}. More recent applications of
the Weinberg
analysis~\cite{Bogn05nuclmat,Bogn06bseries,Rama07WEVPair,Rama13becbcs,Pere14tpep,Srin16PairPNM}
\begin{table*}[t!]
\caption{\label{tab:regulators}
Short- and long-range regulators for the local, semilocal, and nonlocal
potentials of Refs.~\cite{Geze13QMCchi,Geze14long,Epel15improved,Epel15NNn4lo,Carl15sim,Ente17EMn4lo}
with $\rtilde\equiv r/R_0$ and $\ptilde\equiv p/\Lambda$ in the second
and third columns, where $\alpha=(\pi \Gamma(3/4)R_0^3
)^{-1}$ is a normalization constant and $\nu$ is the order in
the chiral expansion. For the EMN potentials, the regulator exponent
$n_2=2$ is applied to the pion exchanges and $n_2=4$ for one-pion
exchange beyond next-to-leading order (NLO).  The highest available chiral order and the
cutoff ranges are given in the fifth column, while the determination
of the $\pi N$ low-energy constants (LECs)/$2 \pi$ regularization and
the fitting protocol are given in the second-to-last and the last
columns, respectively. SFR and DR denote spectral-function and
dimensional regularization, whereas PWA stands for partial-wave analysis.}
\begin{ruledtabular}
\begin{tabular}{l|ccllll} 
& \multicolumn{2}{c}{Regulator functions} & Regulator & Chiral order/ & $\pi N$/ & Fitting protocol \\
& Short & Long & exponent(s) & cutoff range & 2$\pi$ regularization & \\
& (contact) & (pion exchanges) &&& \\[2mm]
\textbf{Local} \\
\cline{1-7} \\
GT+~\cite{Geze14long,Geze13QMCchi} & $\alpha e^{-\rtilde^n}$ & $1 - e^{-\rtilde^n}$ & $n=4$ & \thead{Up to N$^2$LO \\ $R_0=0.9-1.2$~fm} & \thead{Fixed values \\ from Ref.~\cite{Epel05EGMN3LO} \\ SFR} & \thead{Nijmegen PWA~\cite{Stok93PWA}} \\ 

\textbf{Semilocal} \\ 
\cline{1-7} \\ 
EKM~\cite{Epel15improved,Epel15NNn4lo}& $e^{-\ptilde^{n_1}} e^{-\ptildep^{n_1}}$ & $\left( 1-e^{-\rtilde^2}\right)^{n_2}$ & \thead{$n_1=2$\\$n_2=6$} &\thead{Up to N$^4$LO \\ $R_0=0.8-1.2$~fm \\ $\Lambda \approx 493 - 329$~MeV } & \thead{Fixed values~\cite{Epel15improved} \\ DR} & \thead{Nijmegen PWA~\cite{Stok93PWA}} \\ 

\textbf{Nonlocal} \\
\cline{1-7} \\
sim~\cite{Carl15sim} & $e^{-\ptilde^{2n}} e^{-\ptildep^{2n}}$ & $e^{-\ptilde^{2n}} e^{-\ptildep^{2n}}$ & $n=3$ & \thead{Up to N$^2$LO \\ $\Lambda=450-600$~MeV} & \thead{Fitting parameter \\in simultaneous fit \\ SFR} & \thead{Fits to $NN$, $\pi N$, and \\ few-body systems $^{2,3}\text{H}$,$^3\text{He}$} \\
\cline{1-7} \\
EMN~\cite{Ente17EMn4lo} & $e^{-\ptilde^{2n_1}} e^{-\ptildep^{2n_1}}$ & $e^{-\ptilde^{2n_2}} e^{-\ptildep^{2n_2}}$ & \thead{$n_1>\nu/2$ \\ $n_2= 2$ $(4)$} & \thead{Up to N$^4$LO \\$\Lambda=450-550$~MeV} & \thead{Fixed values \\ from Ref.~\cite{Hofe15piNchiral} \\ SFR} & \thead{NN data from 1955-2016~\cite{Berg88psana}} \\
\end{tabular}
\end{ruledtabular}
\end{table*}
provide quantitative insights into how renormalization-group (RG)
techniques act in softening different components of nuclear
interactions and how the effects of potentials are modified at finite
density.

By perturbativeness we mean the order-by-order convergence pattern in
a perturbative many-body expansion (which needs to be distinguished from an order-by-order 
convergence in the chiral EFT expansion). For NN scattering in free space, this expansion is
the Born series.  For many-body systems such as infinite matter and finite
nuclei, this expansion is many-body perturbation theory (MBPT).  While we are
particularly interested in whether MBPT converges and at a practical rate
(e.g., at low-enough order to be tractable), the characterization of
perturbativeness is of more general concern.  For nonperturbative many-body
methods using a basis expansion, the computational resources for convergence
depend strongly on perturbativeness.  It is also relevant for identifying or
justifying reference states such as Hartree-Fock and for motivating
microscopic nuclear density functional theory.

The plan of the paper is as follows.  In
Sec.~\ref{sec:NN_interactions} we characterize three classes of
regularization schemes used in recently formulated chiral NN
interactions and critically compare regulator parameters.  In
Sec.~\ref{subsec:WEA} we review the relevant features of the Weinberg
eigenvalue analysis, illustrating their general behavior in the
complex plane, and document the use of the eigenvalues to approximate
phase shifts for modern interactions.  Eigenvalues at different orders
and in different partial-wave channels are given for the various
chiral NN potentials in Sec.~\ref{sec:results}, highlighting
differences from regulators and features at different orders, which
also depend on the different types of regularization schemes.
Section~\ref{sec:summary} contains our summary and outlook.

\section{NN interactions and regularization}
\label{sec:NN_interactions}

During recent years there has been significant progress in developing
new nuclear forces within chiral EFT (see, e.g.,
Refs.~\cite{Geze13QMCchi,Geze14long,Epel15improved,Epel15NNn4lo,Carl15sim,Ente17EMn4lo}
including also explicit delta resonances in
Refs.~\cite{Piar14Deltas,Piar16DeltaNuc,Ekst17deltasat}). The development of novel
advanced fitting frameworks, the exploration of new regularization
schemes, and the derivation of more systematic ways to estimate
theoretical uncertainties has resulted in new families of interactions
that allow nuclei and nuclear matter to be systematically studied
within \textit{ab initio} frameworks at different orders in the chiral
expansion.

In this section, we briefly summarize properties of these new
interactions to prepare for diagnosing them using the \WEA. In
particular, we focus on three sets of potentials, commonly referred to
as local, nonlocal, and semilocal, which are characterized by
different regularization schemes to separate the long-distance from
the short-distance physics. To be specific, we consider the local
potentials of Refs.~\cite{Geze13QMCchi,Geze14long} by Gezerlis, Tews,
\textit{et al.} (GT+), the semilocal potentials of
Refs.~\cite{Epel15improved,Epel15NNn4lo} by Epelbaum, Krebs, and
Mei{\ss}ner (EKM), the nonlocal potentials of Ref.~\cite{Carl15sim} by
Carlsson, Ekstr\"om \textit{et al.}~(sim), and the nonlocal potentials
of Ref.~\cite{Ente17EMn4lo} by Entem, Machleidt, and Nosyk
(EMN). Table~\ref{tab:regulators} summarizes properties of these
potentials including the specific form of the employed regulators as
well as the available orders in the chiral expansion, the pion-nucleon
($\pi N$) low-energy constants (LECs), the $2 \pi$ regularization, and
the fitting protocols. For more detailed information, we refer to the
given references.

Local interactions use regulators that only depend on the momentum
transfer $\mathbf{q}=\mathbf{p'}-\mathbf{p}$ in momentum space or on
the relative distance $\mathbf{r}$ in coordinate space,
respectively. Here $\mathbf{p}$ and $\mathbf{p'}$ denote the relative
momenta of the initial and final two-body states. The derivation of
local interactions in Refs.~\cite{Geze13QMCchi,Geze14long} opened new
ways for applying nuclear interaction from chiral EFT in
quantum Monte Carlo (QMC)
calculations~\cite{Lynn14QMCln,Tews16QMCPNM,Lynn16QMC3N,Lynn17QMClight}. The
benefits of locally regularizing long-range physics such as the
pion-exchange interactions are discussed in
Ref.~\cite{Epel15improved}. These include the conservation of the
analytical structure of the $T$-matrix close to the pion threshold and
the fact that no spectral function regularization (SFR) is needed in
this regularization approach (see also Ref.~\cite{Geze14long}), with
dimensional regularization (DR) applied in
Ref.~\cite{Epel15improved}. However, for the short-range
couplings the local regularization leads to a mixing of different
partial-wave channels due to the dependence of $\mathbf{q}$ on the angle
$\cos \theta_{\mathbf{p} \mathbf{p}'}$. As a consequence,
$S$-wave short-range contact interactions generally induce
nonvanishing contributions in higher partial waves after
regularization~\cite{Geze14long}, whereas for nonlocal regulators,
which only depend on the magnitude of the relative momenta $p$ and
$p'$, such short-range interactions remain restricted to only
$S$ waves. This leads in particular to technical simplifications since
different partial-wave channels can be fitted independently.

The semilocal EKM interactions~\cite{Epel15improved,Epel15NNn4lo}
combine the conceptual advantages of locally regularized long-range
interactions with technical benefits of nonlocal short-range
interactions. In practice, the regularization of the long-range parts
is formulated in coordinate space and is characterized by a cutoff
scale $R_0$, whereas the short-range regularization is performed in
momentum space which involves a cutoff scale $\Lambda$. Physically, it
is a natural assumption that these two scales should be related. In
Ref.~\cite{Epel15improved}, a mapping between the two scales was
motivated by considering the Fourier transforms of Gaussians, which
leads to the relation
\begin{equation} 
\label{eq:Lambda_semilocal}
\Lambda(R_0) = \frac{2}{R_0} \,.
\end{equation}
In Ref.~\cite{Geze14long}, a cutoff mapping between momentum and
coordinate space was suggested by relating the integral over the
Fourier-transformed short-range regulator function (see
Table~\ref{tab:regulators}),
\begin{equation}
\label{eq:local_short_mom}
f_{\rm local}(q^2,R_0) = \int \text{d} \vec{r} \ \alpha e^{-(r/R_0)^4} e^{-i\vec{q} \cdot \vec{r}}\,,
\end{equation}
with the integral over a sharp momentum cutoff:
\begin{equation}
\label{eq:Lambda_local}
\int \text{d} \vec{q} \: f_{\rm local}(q^2,R_0) = \int \text{d} \vec{q} \: \theta(\Lambda-|\mathbf{q}|) \,.
\end{equation}

\begin{figure}[t!]
\includegraphics[page=1,scale=1.0,clip]{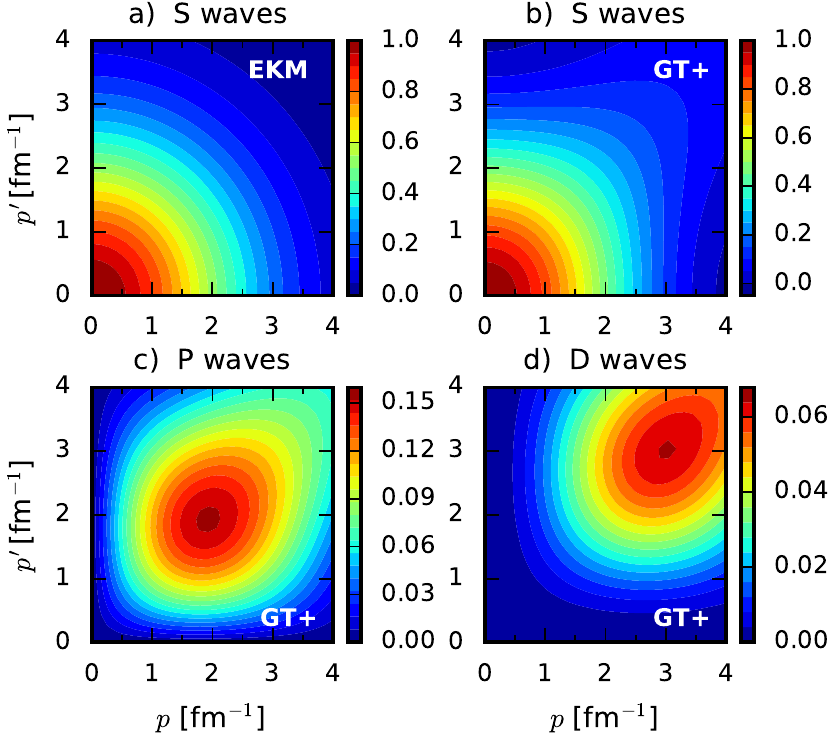}
\caption{\label{fig:reg_short_PW}(Color online)
Contour plot for the short-range regulator of the GT+ and EKM
potentials. (a) The nonlocal EKM regulator, which is independent of the
angular momentum, is plotted for $\Lambda = 493$~MeV, while
the local regulator, which depends on the partial wave, is shown in
the (b) $S$, (c) $P$, and (d) $D$ waves for the cutoff
$R_0=1.2$~fm. We find good agreement in
the $S$ waves for the cutoff combination $R_0^\text{GT+}=1.2$~fm and
$R_0^\text{EKM}=0.8$~fm, assuming Eq.~\eqref{eq:Lambda_semilocal}. A
least-squares minimization shows that the regulators in the $S$ waves
are most comparable for $R_0^{\text{EKM}} = 0.85$~fm.}
\end{figure}

Obviously, there is no universal way to relate the coordinate and
momentum space cutoff scales. By comparing the numerical values for
$\Lambda$ resulting from relations (\ref{eq:Lambda_semilocal}) and
(\ref{eq:Lambda_local}) we obtain quite different numbers: for $R_0 =
0.8$~fm we get $\Lambda = 493$ and $614$~MeV,
whereas for $R_0 = 1.2$~fm, $\Lambda = 329$ and
$409$~MeV, respectively. In Fig.~\ref{fig:reg_short_PW} we show a contour plot of
the semilocal short-range regulator with $R_0^{\text{EKM}}=0.8$~fm, i.e., $\Lambda
= 493$~MeV in the $S$ waves, and the Fourier transform of the local
short-range regulator with $R_0^{\text{GT+}}=1.2$~fm in the $S$, $P$, and
$D$ waves.  We find good agreement in the $S$ waves for this chosen
cutoff combination, with a least-squares minimization indicating best
agreement for $R_0 = 0.85$~fm for the semilocal potential.  However,
we observe in general a quite different behavior for the nonlocal versus the
angular-dependent local regulator in momentum space, where the latter
does not cut off contributions with $\vec{p}=\vec{p}'$. Furthermore, the $q^2$-dependent
contacts at NLO and beyond with $p \neq p'$ are cut off much slower by
the local regulator.  This shows that the comparison of the numerical
values of $R_0$ alone can be quite misleading due to the different
regulator forms for different interactions.

\begin{table}[t!]
\caption{\label{tab:half_val}
Distance $r^*$ where the long-range regulator function takes the value
$f_\text{long}(r^*, R_0)=1/2$ for the GT+ (middle column) and EKM (right
column) regulator functions; see Table~\ref{tab:regulators}. Results are shown
for a cutoff range of $R_0=0.8$--$1.2$~fm. We find best agreement for the cutoff
combination $R_0^\text{GT+}=1.2$~fm and $R_0^\text{EKM}=0.8$~fm.}
\begin{ruledtabular}
\begin{tabular}{ccc} 
$R_0$ [fm] & $r^*_{\text{GT+}}$ [fm] & $r^*_{\text{EKM}}$ [fm] \\
\hline
0.8 & 0.73 & 1.19 \\
0.9 & 0.82 & 1.34 \\
1.0 & 0.91 & 1.49 \\
1.1 & 1.00 & 1.64 \\
1.2 & 1.10 & 1.79 \\
\end{tabular}
\end{ruledtabular}
\end{table}

\begin{figure}[t!]
\includegraphics[page=1,scale=1.0,clip]{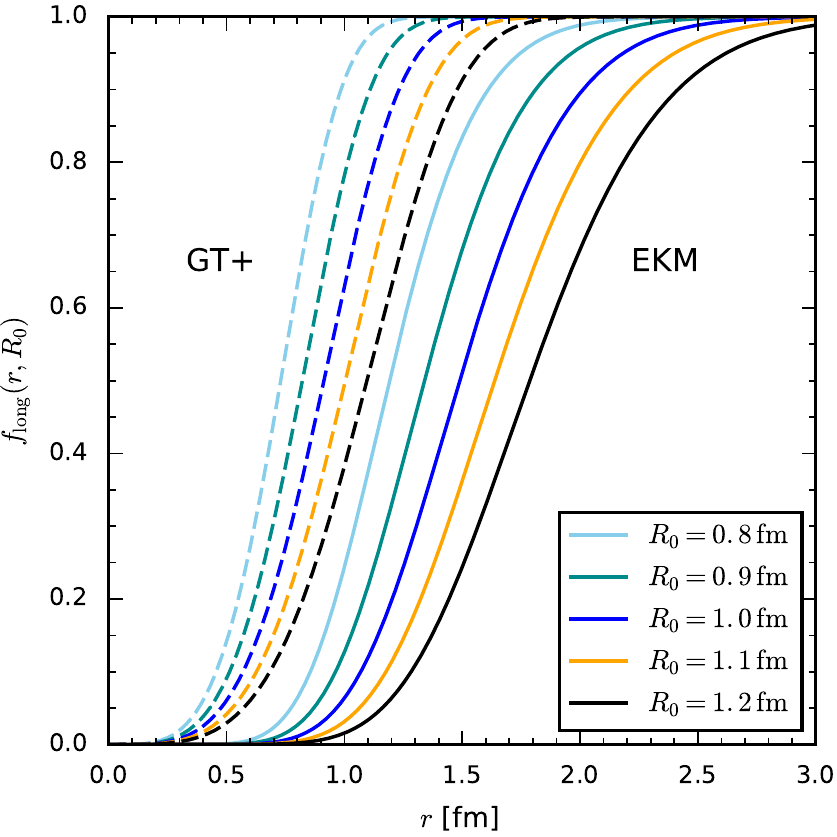}
\caption{\label{fig:reg_long_comparison}(Color online) 
Plot of the long-range regulator functions for the GT+ (dashed) and
EKM (solid) potentials with cutoffs $R_0=0.8$--$1.2$~fm (see
Table~\ref{tab:regulators}). The regulators corresponding to
$R_0^\text{GT+} = 1.2$~fm (dashed black line) and $R_0^\text{EKM} =
0.8$~fm (solid light blue line) lead to the best agreement.}
\end{figure}

We can confirm this observation also for the long-range part of the
regulators. In Table~\ref{tab:half_val} we show the distance $r^*$,
where $f_\text{long}(r^*, R_0)=1/2$ for the cutoff range
$R_0=0.8$--$1.2$~fm.  Similarly to the short-range part of the
regulators, we find good agreement for $R_0^\text{GT+}=1.2~\text{fm}$
and $R_0^\text{EKM}=0.8~\text{fm}$.
As shown in Fig.~\ref{fig:reg_long_comparison}, the regulator functions agree well over the entire range of
distances. Therefore, it is natural to expect that the Weinberg
eigenvalue analysis will provide similar results for the long-range part
for this cutoff combination of these two interactions, which we focus on in Sec.~\ref{sec:results}.

\section{Weinberg eigenvalue analysis}
\label{subsec:WEA}

The Weinberg eigenvalue analysis is a powerful tool to quantify the
perturbativeness of nuclear interactions. Perturbativeness is of great
importance for most of the many-body frameworks presently used in
nuclear physics.  On the one hand, the tractability of MBPT directly
relies on the rapid convergence of the perturbation series through
suppression of higher-order corrections, because otherwise the number
of diagrams increases too fast with successive orders.  To date, MBPT
has been applied to the calculation of the equation of state of
infinite nuclear matter up to third order (see, e.g.,
Refs.~\cite{Hebe11fits,Tews13N3LO,Cora14nmat,Well16DivAsym,Dris16asym,Dris16nmatt,Holt16eos3pt}) and recently to fourth order~\cite{Dris17Mcshort}.
In addition, MBPT has been applied to the derivation of valence-space
Hamiltonians for open-shell nuclei (see, e.g.,
Refs.~\cite{Hebe15ARNPS,Simo16unc}) and recently to the calculation of
ground-state energies of closed-shell nuclei~\cite{Tich16HFMBPT}.

On the other hand, perturbativeness also plays a key role for
inherently nonperturbative many-body frameworks that are based on
basis expansions such as the no-core-shell model~\cite{Barr13PPNP},
coupled-cluster theory~\cite{Hage14rev}, in-medium similarity
renormalization group~\cite{Herg16PR}, and the self-consistent Green's
function method~\cite{Carb13nm,Soma14GGF2N3N}. For these frameworks,
strongly nonperturbative interactions typically require a
prohibitively large number of basis states and prevent a reliable
extraction of converged results.  In recent years, RG methods have
been developed in order to improve the perturbativeness of nuclear
interactions.  However, such RG transformations can generally only be
performed approximately and thus lead to additional uncertainties in
many-body calculations~\cite{Bogn10PPNP,Furn13RPP}.

We review here briefly the most important aspects of the Weinberg
eigenvalue analysis and refer to Ref.~\cite{Wein63quasip} for more
detailed discussions.  To motivate the concept, we consider for
simplicity the Lippmann-Schwinger equation for the free-space
$T$-matrix in the center-of-mass frame,
\begin{subequations} 
\label{eq:T_mat_eqs}
\begin{align}
T(W) &= V + V G_0(W) T(W)  \label{eq:T_mat_eq} \\
&= \sum \limits_{n=0}^\infty V \left( G_0(W) V \right)^n \,, 
\label{eq:T_mat_eq_born}
\end{align}
\end{subequations}
with the free propagator $G_0(W)=(W - H_0)^{-1}$, the kinetic energy
$H_0 = p^2/m$, where $m$ is the averaged nucleon mass, and $W$ is the
complex energy.

Iteration of the Born series~\eqref{eq:T_mat_eq_born} may converge to a
self-consistent solution. Due to nonperturbative sources, however, the
convergence is by no means guaranteed; e.g., bound states are poles of the
$T$-matrix, which render the expansion naturally divergent. To study
convergence and the efficiency of perturbation theory, Weinberg analyzed 
the eigenvalues of the operator $G_0(W)V$,
\begin{equation}
\label{eq:weinb_eig_rel}
G_0(W) V \ket{\Psi_\nu(W)} = \eta_\nu(W) \ket{\Psi_\nu(W)} \, .
\end{equation}
The so-called Weinberg eigenvalues $\eta_\nu(W)$ are defined in the
complex energy plane cut along the positive real axis and form a
discrete set for any value of $W$. In the following, we take
$W=E+i\varepsilon$ for positive energies.

Making use of the eigenvalue relation~\eqref{eq:weinb_eig_rel}, the
Born series expansion~\eqref{eq:T_mat_eq_born} is a geometric series
which converges if and only if all eigenvalues lie within the unit
circle in the complex plane, i.e., $|\eta_\nu(W)| < 1$. The largest
eigenvalue sets the rate of convergence, if at all, where overall
smaller magnitudes imply faster convergence. When $|\eta_\nu(W)| > 1$,
the precise magnitudes of the Weinberg eigenvalues still have a
dramatic impact on the convergence in a nonperturbative many-body
method.

We summarize here several definitions as well as selected properties of
$\eta_\nu(W)$ relevant for this paper. Rewriting the eigenvalue
relation~\eqref{eq:weinb_eig_rel} as a modified Schr{\"o}dinger
equation~\cite{Wein63quasip},
\begin{equation} \label{eq:mod_schroed}
\left( H_0 + \frac{V}{\eta_\nu(W)} \right) \ket{\Psi_\nu(W)} 
= W \ket{\Psi_\nu(W)} \,,
\end{equation}
allows intuitively a physical interpretation: the eigenvalue is
effectively an energy-depending coupling $\eta_\nu^{-1}(W)$ which
rescales the interaction.  Following the original discussion by
Weinberg, real bound states of the potential having $W = E<0$ (e.g., for
the deuteron, $E=-2.223$~MeV) correspond to $\eta_\nu(E)=1$. The
modified Schr{\"o}dinger equation corresponds to the physical one in
this case. More generally, even though the original potential does not
support a bound state with binding energy $E<0$, a scaled interaction
$\eta_\nu^{-1}(E) V$ would have a bound state at $E$.

A purely attractive potential has only positive eigenvalues for $E<0$. However, a purely repulsive potential cannot have a bound-state
solution of the Schr{\"o}dinger equation, which naively seems to imply that the modified
Schr{\"o}dinger equation \eqref{eq:mod_schroed} has no solutions.
However, \eqref{eq:mod_schroed}  may have a solution for a sign-flipped
interaction $\eta_\nu^{-1}(E) V$ in which the Weinberg eigenvalue is
negative.  Therefore, it is convention that a positive (negative)
eigenvalue is referred to as an attractive (repulsive) eigenvalue.  

In the case of positive energies ($E > 0$) for $W=E+i\varepsilon$ with
$\varepsilon \to 0$, the modified Schr{\"o}dinger equation has complex
energy eigenvalues, leading to complex Weinberg eigenvalues. Thus, we
obtain complex (real) eigenvalues for positive (negative) energies
$E$.  The same definition of attractive and repulsive as before
applies to the imaginary part of the eigenvalues for positive
energies, which is motivated by analytic continuation from the
solution along the negative real axis.  In general, both attractive
and repulsive eigenvalues occur for a nuclear potential.

\begin{figure}[t!]
\includegraphics[page=1,scale=1.0,clip]{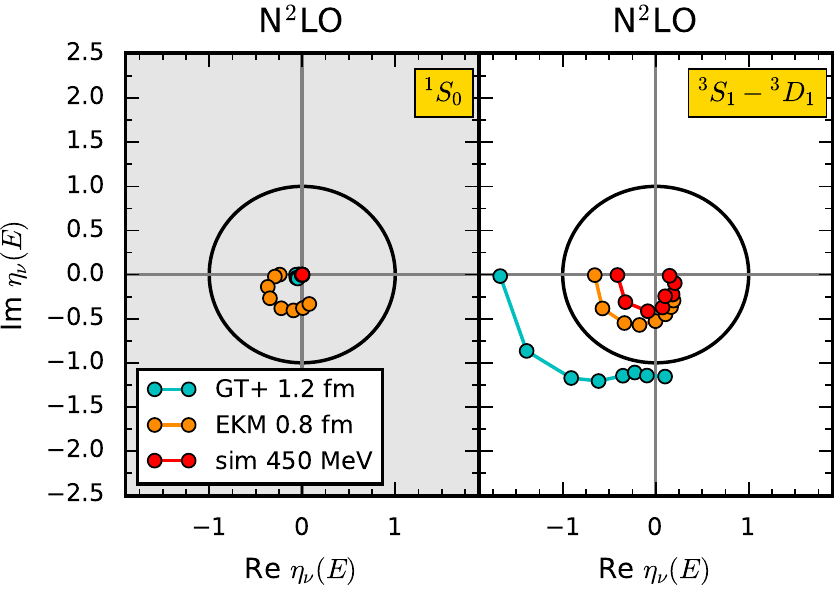}
\caption{\label{fig:rep_loc_semi_sim_c_plane}(Color online)
Repulsive Weinberg eigenvalues for the N$^2$LO NN potentials GT+ 
$1.2$~fm, EKM $0.8$~fm, and sim $450$~MeV ($T_{\rm rel}=290$~MeV) as
trajectories of energy in the complex plane, starting on the negative
real axes and evolving counterclockwise. We show results for energies
$E=0, 25, 66, 100, 150, 200, 250,$ and $300$~MeV as circles in the \oneSzero (left
panel) and \threeSDone channels (right panel).}
\end{figure}

\begin{figure}[t!]
\includegraphics[page=1,scale=1.0,clip]{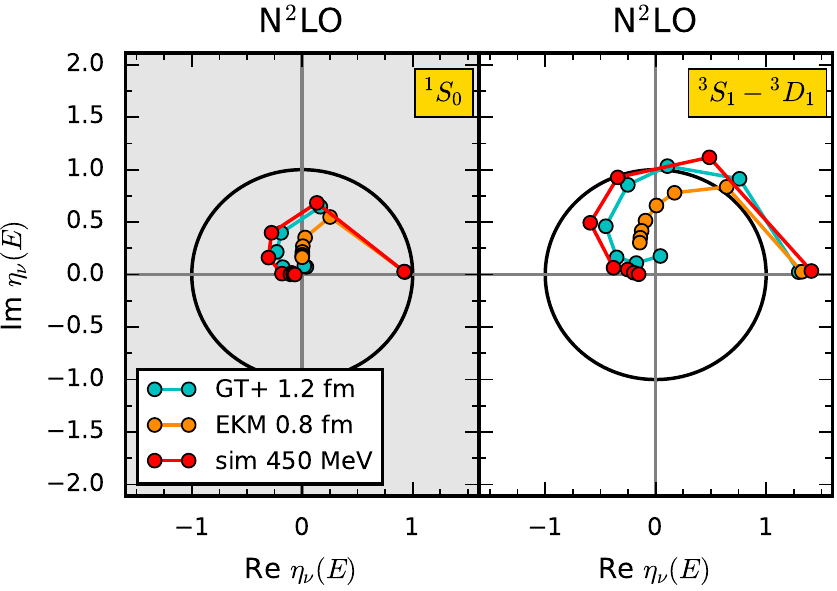}
\caption{\label{fig:attr_loc_semi_sim_c_plane}(Color online)
Attractive Weinberg eigenvalues for the N$^2$LO NN potentials GT+ 
$1.2$~fm, EKM $0.8$~fm, and sim $450$~MeV ($T_{\rm rel}=290$~MeV) as
trajectories of energy in the complex plane, starting on the positive
real axes and evolving counterclockwise. We show results for energies
$E=0, 25, 66, 100, 150, 200, 250,$ and $300$~MeV as circles in the \oneSzero (left
panel) and \threeSDone channels (right panel). The (nearly and shallow)
bound states close to $E=0$ are indicated by eigenvalues slightly
smaller or larger than $1$ in the uncoupled or coupled channel,
respectively.}
\end{figure}

We illustrate the behavior of repulsive and attractive Weinberg
eigenvalues in the complex plane for positive energies $E=0-300$~MeV
in Figs.~\ref{fig:rep_loc_semi_sim_c_plane} and
\ref{fig:attr_loc_semi_sim_c_plane}, respectively, in the \oneSzero
and \threeSDone channels for a set of three different potentials by
taking the limit $\varepsilon \to 0$ of $\eta_\nu(E+i\varepsilon)$.
The trajectories start on the real axis and evolve counterclockwise
with increasing energy.  Nearly (or shallow) bound states are
represented by attractive eigenvalues with magnitudes close to unity
for $E=0$. The deuteron binding energy can be determined by the
intersection of the trajectory in the \threeSDone channel and the unit
circle when lowering the energy $E<0$. Since the attractive
eigenvalues are typically dominated by (nearly or shallow) bound
states in the two $S$-wave channels, we discuss in the present paper
mainly repulsive eigenvalues.

We also briefly give some details of the calculation. In practice, it
is convenient to solve the eigenvalue
relation~\eqref{eq:weinb_eig_rel} in a partial-wave representation
because $G_0V(W)$ is block diagonal in the partial-wave quantum
numbers $(LS)JT$,
\begin{equation}
\label{eq:wein_block_mat}
\bordermatrix{& $\oneSzero$ & $\threeSone$ & $\threeDone$ & $\onePone$ & \ldots \quad \cr
$\oneSzero$ & \blacksquare & 0 & 0 & 0 & \ldots \quad \cr
$\threeSone$ & 0 & \blacksquare & \blacksquare & 0 & \ldots \quad \cr
$\threeDone$ & 0 & \blacksquare & \blacksquare & 0 & \ldots \quad \cr
$\onePone$ & 0 & 0 & 0 & \blacksquare & \cr
\vdots & \vdots & \vdots & \vdots & \vdots & \ddots \quad} \,,
\end{equation}
where $L$ denotes the angular momentum, $S$ the two-body spin, $J$ the
total angular momentum, and $T$ the two-body isospin.  This allows one to
separately diagonalize blocks of given $S,J,$ and $T$ (the $\blacksquare$ in
Eq.~\eqref{eq:wein_block_mat}),
\begin{multline} 
\label{eq:weinb_eig_rel_pwd}
\frac{2}{\pi}\sum \limits_{L,L'} \int \text{d}k' \, \frac{k'^2 m V_{L L' S}^{JT}(k,k')}{k^2_0-k'^2 + i \varepsilon} \braket{k'(L'S)JT|\Psi_\nu(W)} \\
= \eta_\nu(W) \sum \limits_L \braket{k(LS)JT|\Psi_\nu(W)} \,,
\end{multline}
where different $L$ values may be coupled due to the potential ($k_0^2+ i\varepsilon
= mW$). For coupled channels, we have $L,L'= |J \pm 1|$, whereas in
uncoupled channels $L=L'$. The main discussion of this paper is based
on the free propagator and on the neutron-proton (np) channel but
isospin-symmetry breaking is usually small. Hence, we have dropped the
index $M_T=0$ for simplicity.

In the case of negative energies (i.e., purely imaginary $k_0$), poles do
not occur and we can take $\varepsilon=0$. Technically, we then solve
the eigenvalue problem on a well-suited Gaussian quadrature momentum
grid to ensure numerical convergence. After performing the standard
substitution $\int \mathrm{dp} \to \sum_{i=1}^{N_p}$, the left-hand
side of the eigenvalue problem~\eqref{eq:weinb_eig_rel_pwd} can be
written as a matrix. The basis vectors have a size of $N_p$ ($2N_p$)
in an uncoupled (coupled) channel. 

For the positive energies, however, one has to carefully take into
account the pole in Eq.~\eqref{eq:weinb_eig_rel_pwd} at $k = k_0$. In
that case, we make use of the Sokhotski-Plemelj theorem for a real,
continuous function $f(k)$,
\begin{equation} \label{eq:sokhotski_plemelj}
\frac{f(k)}{k-(k_0\pm i \varepsilon)}= \mathcal{P} \frac{f(k)}{k-k_0}\pm i \pi \delta(k-k_0)f(k) \,,
\end{equation}
with the Cauchy principal value $\mathcal{P}$, and integrate explicitly over
the singularity. Following Ref.~\cite{Brow69MinRel}, we convert the
principal-value integral into a standard integral by adding
\begin{equation}
-g(k_0) \, \mathcal{P} \int_0^\infty \frac{\text{d}k}{k^2-k_0^2} = 0
\end{equation}
to Eq.~\eqref{eq:weinb_eig_rel_pwd} in order to make the integral
well behaved, i.e.,
\begin{equation}\label{eq:rewrite_pv_int}
\mathcal{P} \int_0^{\infty} \text{d}k \, \frac{g(k)}{k^2-k_0^2} = 
\int_0^{\infty} \text{d}k \, \frac{g(k)-g(k_0)}{k^2-k_0^2} \,,
\end{equation}
where we define $f(k) = g(k)/(k+k_0)$. To evaluate numerically the
integral on the right-hand side of Eq.~\eqref{eq:rewrite_pv_int}, it
is crucial to split the integral at some sufficiently large
$p_\text{max} > k_0$ such that $f(k)$ is known to vanish for all $p >
p_\text{max}$. Because of the regularization of the potential, it is
usually straightforward to find a suitable value for
$p_\text{max}$. The advantage of this procedure is that the remaining
integral of the form
\begin{equation}
\int_{p_\text{max}}^\infty \frac{\text{d}k}{k^2-k_0^2} = \frac{1}{k_0}
\operatorname{artanh} \left( \frac{k_0}{p_\text{max}} \right) 
\end{equation}
no longer has a pole because of $p_\text{max} > k_0$, and can be
evaluated analytically. We have carefully checked the numerical
stability of this method, in particular the subtraction in
Eq.~\eqref{eq:rewrite_pv_int}. The subtracted pole as well as the
additional constant term in Eq.~\eqref{eq:sokhotski_plemelj} are taken
care of by enlarging the basis vector by one for each $L$ component,
so the matrix to be diagonalized is of rank $N_p+1$ $(2N_p+2)$ for an
uncoupled (coupled) channel.

\begin{figure}[t!]
\includegraphics[page=1,scale=1.0,clip]{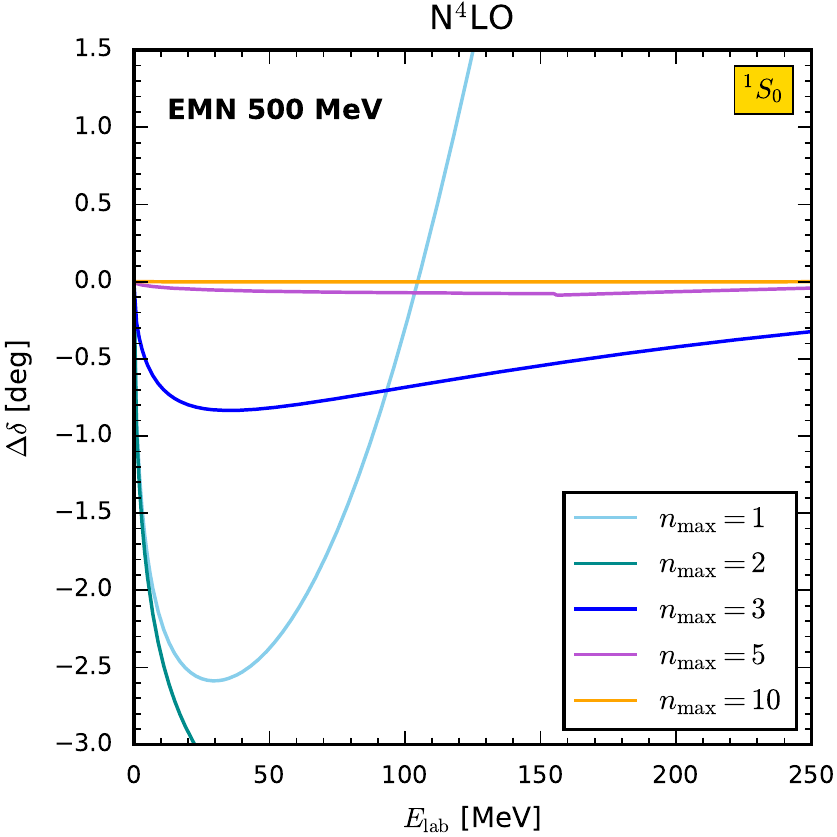}
\caption{\label{fig:phase_shifts_by_WE}(Color online)
Convergence pattern of \oneSzero phase shifts calculated using the
largest Weinberg eigenvalues and different truncations $n \leqslant
n_\text{max}$ in expansion~\eqref{eq:wein_bare_phaseshifts}. The results are based on
the 500~MeV EMN potential at N$^4$LO, but the other potentials and
channels show a similar behavior, and the phase shifts are plotted as a
function of $E_\text{lab}=2E$. Note that for $n_\text{max}=1$,
we restrict the sum to the largest attractive (instead of the overall
largest) eigenvalue to avoid a discontinuity that happens because the
trajectories of attractive and repulsive eigenvalues are crossing each
other.}
\end{figure}

Finally, we review an intriguing feature of the Weinberg analysis. Weinberg
showed in Section~VI of Ref.~\cite{Wein63quasip} that the eigenvalues and
phase shifts in an uncoupled channel $(LS)JT$ are related by 
\begin{align}
\delta_{LS}^{JT}(E) &= \sum \limits_{\nu=1}^\infty \delta_{\nu}(E) \,, 
\label{eq:wein_bare_phaseshifts}
\intertext{with the so-called elemental phase shifts defined as}
\delta_{\nu}(E) &\equiv -\operatorname{arg} \left( 1-\eta_{\nu}(E + i\varepsilon) \right) \,, \label{eq:wein_elem_phaseshifts}
\end{align}
where the $\eta_{\nu}$ are solutions to
Eq.~\eqref{eq:weinb_eig_rel_pwd} for the uncoupled channel. For
coupled channels, Eq.~\eqref{eq:wein_bare_phaseshifts} leads to the
sum of the partial phase shifts, $\delta_{L-1S}^{JT} +
\delta_{L+1S}^{JT}$, which is independent of a particular phase-shift
convention. Repulsive (attractive) eigenvalues lead to elemental
phase shifts in $\left[-\pi,0\right]$ ($\left[0,\pi\right]$)
resulting, as expected for purely repulsive (attractive) interactions,
in negative (positive) phase shifts.

Weinberg already observed that Eq.~\eqref{eq:wein_bare_phaseshifts}
usually converges rapidly, taking into
account only a few terms. Consequently, there can only be a few
eigenvalues with significant magnitudes. We find a similar convergence
pattern also for our representative set of modern chiral
potentials. In Fig.~\ref{fig:phase_shifts_by_WE}, we show the
residuals
\begin{equation}
\Delta\delta_{LS}^{JT}(E) = \sum_{\nu=1}^{n_\text{max}}\delta_{\nu}(E) - \delta_{LS}^{JT}(E) \,,
\end{equation}
evaluated for several truncations $n_\text{max}$. The results in
Fig.~\ref{fig:phase_shifts_by_WE} are shown for the
500~MeV~EMN  potential at N$^4$LO in the \oneSzero channel;
however, the other potentials and channels discussed in this paper behave
similarly. The reference phase shifts $\delta_{LS}^{JT}(E)$ result from the
on-shell $T$ matrix as obtained in a nonperturbative calculation by inverting
Eq.~\eqref{eq:T_mat_eq}. The converged phase shifts are very well reproduced
for $n_\text{max} \sim 5 -10$.

\section{Results}
\label{sec:results}

\begin{figure*}[p!]
\includegraphics[page=1,width=\textwidth,clip]{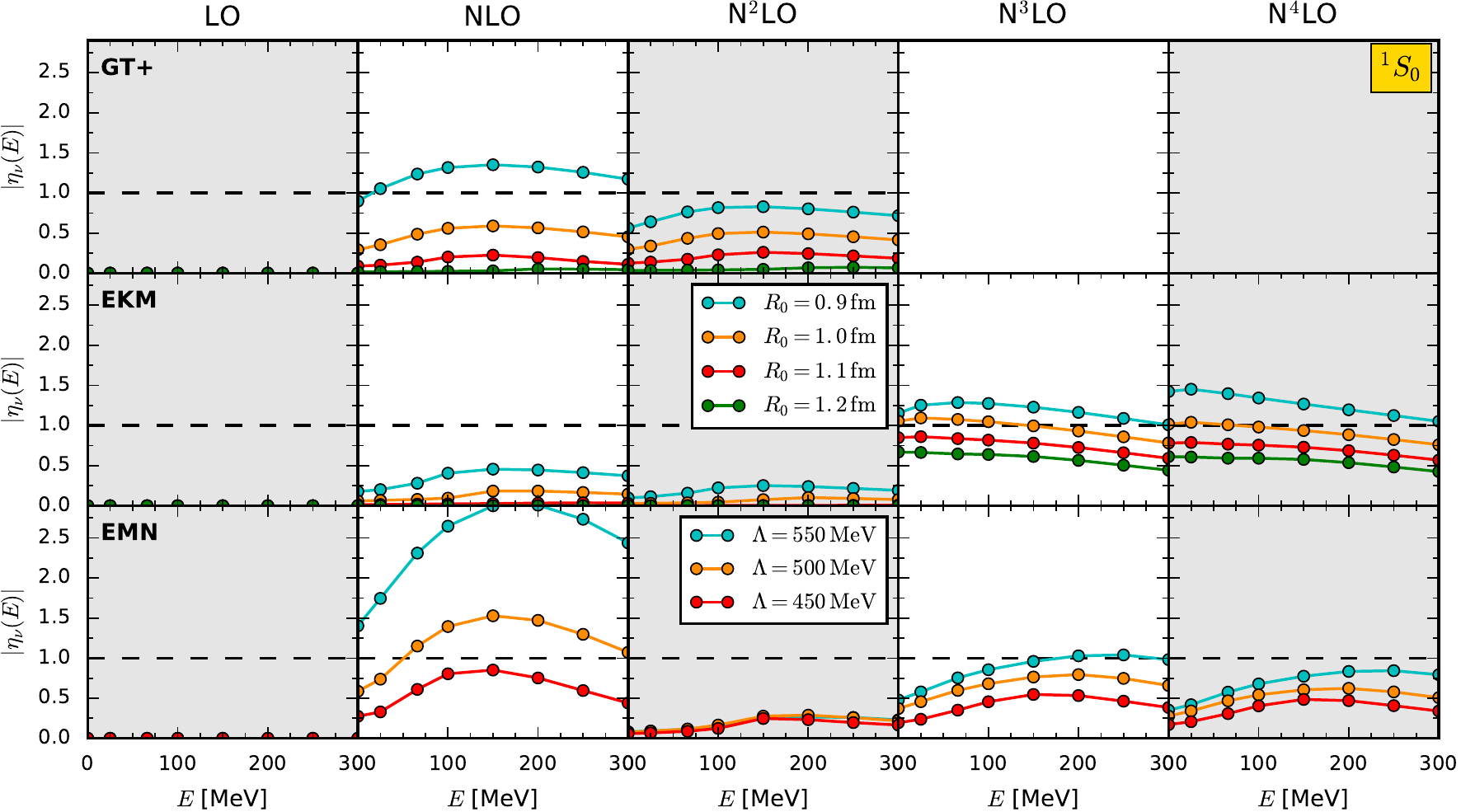}
\caption{\label{fig:rep_WE_loc_semiloc_EM_new_1s0}(Color online)
Magnitude of the repulsive Weinberg eigenvalues for the GT+ (first row), EKM
(middle row), and  EMN potentials (bottom row) as a function of energy $E=0, 25,
66, 100, 150, 200, 250,$ and $300$~MeV in the \oneSzero channel up to the highest chiral
order available, respectively. We show results for coordinate-space cutoffs
$R_0=0.9-1.2$~fm for the GT+ and EKM potentials, as well as for momentum-space
cutoffs $\Lambda=450-550$~MeV for the EMN potential.}
\bigskip
\includegraphics[page=1,width=\textwidth,clip]{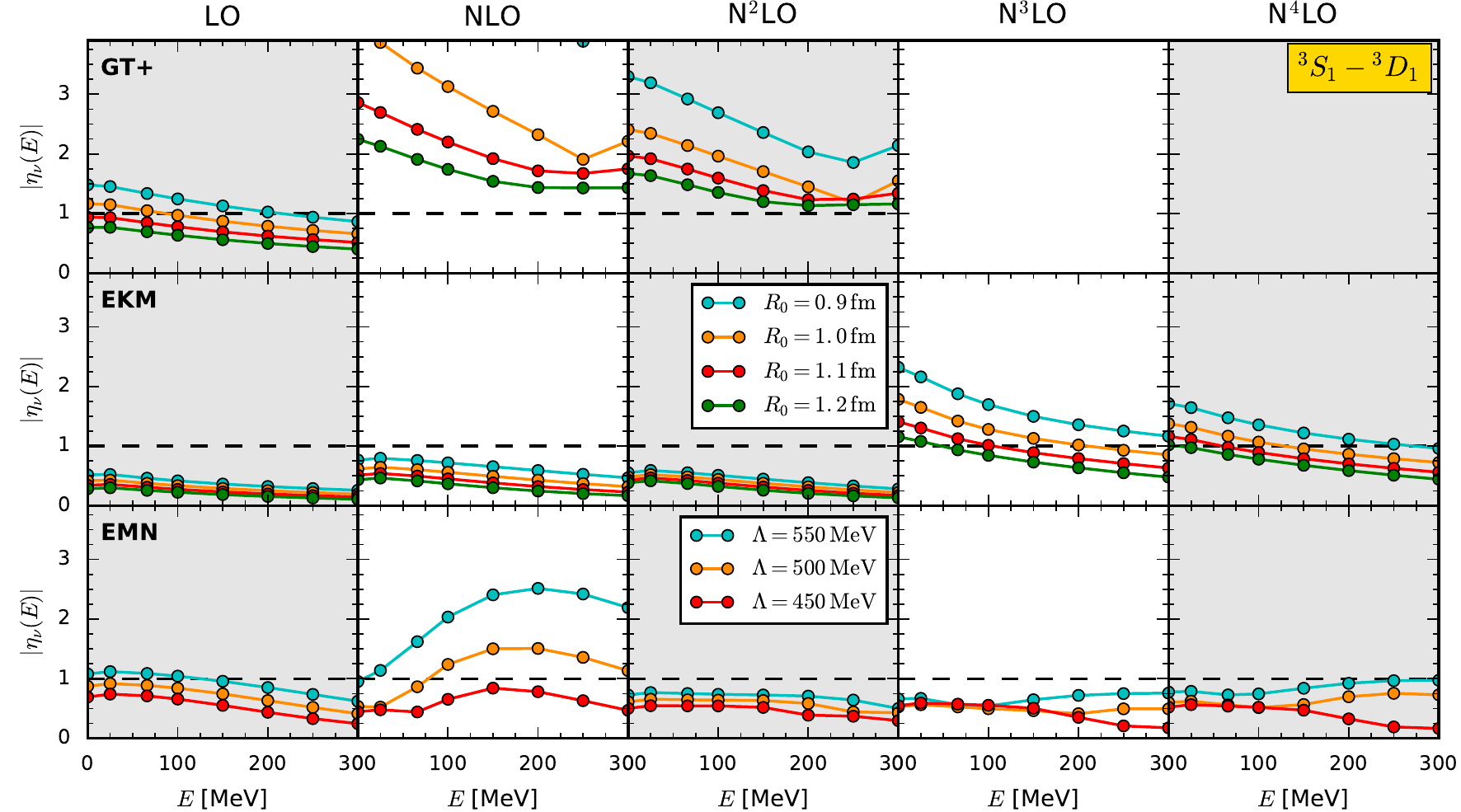}
\caption{\label{fig:rep_WE_loc_semiloc_EM_new_3sd1}(Color online)
Same as Fig.~\ref{fig:rep_WE_loc_semiloc_EM_new_1s0} but for the \threeSDone
channel. Notice that the Weinberg eigenvalues are above the scale for the NLO
NN potential GT+ $0.9$~fm, as we use the same plot range for all panels for 
better comparison.}
\end{figure*}

In this section, we apply the Weinberg eigenvalue analysis to the
recent local, semilocal, and nonlocal chiral potentials in different
partial waves. We investigate and compare characteristic features of
each potential order by order and exploit the regulator comparison of
Sec.~\ref{sec:NN_interactions} for the local and semilocal
potentials. The repulsive eigenvalues manifest the differences between
the various potentials, so we focus our analysis on them, but also
illustrate the common trends of attractive eigenvalues. At the end we
revisit the question of whether distinct but phase-equivalent initial
potentials flow to the same low-momentum form under the similarity RG.

We start with the \oneSzero and coupled \threeSDone channels, as they
are most important for low-energy physics, and then extend the
discussion to higher partial waves. In
Figs.~\ref{fig:rep_WE_loc_semiloc_EM_new_1s0}
and~\ref{fig:rep_WE_loc_semiloc_EM_new_3sd1}, we show the magnitude of
the $S$-wave repulsive eigenvalues as a function of energy from leading order (LO) up to the
highest order available, respectively, for the local GT+, semilocal EKM, and nonlocal EMN potentials in each row with various cutoffs. The
dotted black line denotes where the Born series expansion diverges,
corresponding to the unit circle in
Figs.~\ref{fig:rep_loc_semi_sim_c_plane}
and~\ref{fig:attr_loc_semi_sim_c_plane}.  For the GT+ potential we use
the SFR cutoff $\tilde{\Lambda}=1000$~MeV. From these figures, we
observe the following:
\begin{itemize}

  \item In the \oneSzero channel, all three LO potentials are purely
    attractive and so the repulsive eigenvalues are zero.  In
    contrast, the corresponding eigenvalues in the \threeSDone channel
    are nonzero and show significant differences, with the EKM
    potentials softer than GT+ and EMN.

  \item At NLO we find nonvanishing repulsive eigenvalues, large in
    magnitude for the GT+ potential and even larger for the EMN
    potential in the \oneSzero channel. In the \threeSDone channel we
    observe magnitudes up to 8 for the GT+ 0.9~fm potential and up to
    $2.5$ for the EMN~550~MeV potential, while eigenvalues are below
    1 for the EKM potential in both channels.

  \item Going from NLO to N$^2$LO leads to reduced eigenvalues
    uniformly, with EMN in particular going from nonperturbative for
    the larger $\Lambda$ values to perturbative.
  
  \item The eigenvalues for the EKM and EMN potentials in the
    \oneSzero channel jump upwards at N$^3$LO and stay equally large
    in magnitude at N$^4$LO. In the \threeSDone channel, the eigenvalues for
    the EKM potential again increase at N$^3$LO and N$^4$LO, whereas for the
    EMN potential we observe essentially no change in magnitude but an
    increased spread in $\Lambda$ for higher energies.  Enhanced repulsive
    eigenvalues at N$^3$LO were discussed in Ref.~\cite{Bogn06bseries}
    due to the sub-sub-leading two-pion exchange as a new nonperturbative
    source entering at N$^3$LO. It is interesting to note that these
    jumps in the eigenvalues are also manifested in the form of large energy
    changes of the triton binding energy~\cite{Bind16SoANN,Ente17EMn4lo} based on these two-body
    interactions~\cite{Mach17privcom}.

\end{itemize}

\begin{figure}[t!]
\includegraphics[page=1,scale=0.9,clip]{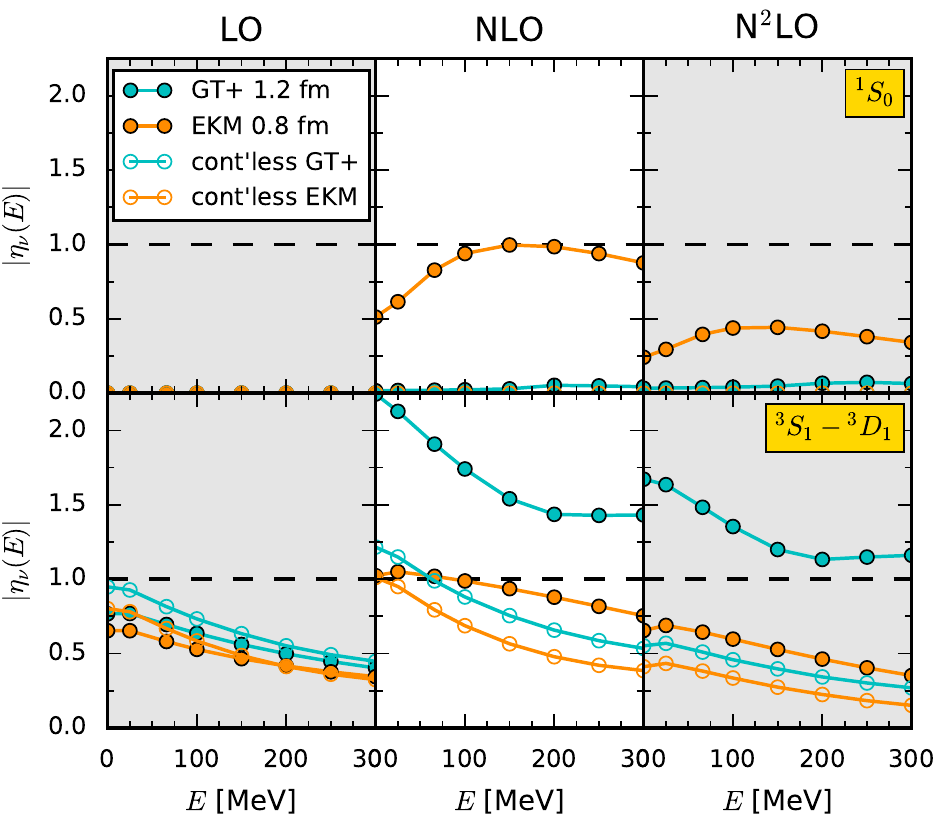}
\caption{\label{fig:contactless_local_semilocal_fixed_R0}(Color online)
Magnitude of the repulsive Weinberg eigenvalues for the GT+ and EKM 
potential for the fixed cutoff combination derived in
Sec.~\ref{sec:NN_interactions}, at LO, NLO, and N$^2$LO. We show
results for the full potential (solid circles) in contrast to the
potential without contacts (open circles) in the \oneSzero (upper
panel) and \threeSDone channels (lower panel). The eigenvalues for the
contactless potential are in fair agreement for the cutoff combination
$R_0^\text{GT+}=1.2~\text{fm}$ and $R_0^\text{EKM}=0.8~\text{fm}$ at
all orders and in both channels.}
\end{figure}

All potentials at all orders get softened for larger coordinate-space cutoffs
or smaller momentum-space cutoffs, respectively, resulting in less repulsion
and therefore smaller repulsive eigenvalues. 
In general, the larger
eigenvalues of the local GT+ potentials indicate that it is less
perturbative than the semilocal or nonlocal potentials.
This observation is consistent with past studies of local versus nonlocal
one-boson-exchange potentials~\cite{Mach95nonlocal}.
However, as discussed in Sec.~\ref{sec:NN_interactions}, a direct
comparison of the local GT+ and semilocal EKM potentials with the same
regulator parameter $R_0$ is misleading because of the differing forms of the
regulator functions. We identified comparable cutoff values, but good
agreement for eigenvalues of the corresponding full potentials is only seen at
LO. In Fig.~\ref{fig:contactless_local_semilocal_fixed_R0} we compare the full
and contactless potentials to shed light on the deviations. In this context,
contactless means all contacts up to the given chiral order are set to zero.
We find fair agreement for eigenvalues of the contactless potentials in both
channels, even at NLO and N$^2$LO. Thus we conclude that the different
inclusion of the momentum-dependent short-range couplings (for local, and
semilocal or nonlocal) at NLO and beyond lead to the differences in eigenvalues.

\begin{figure}[t!]
\includegraphics[]{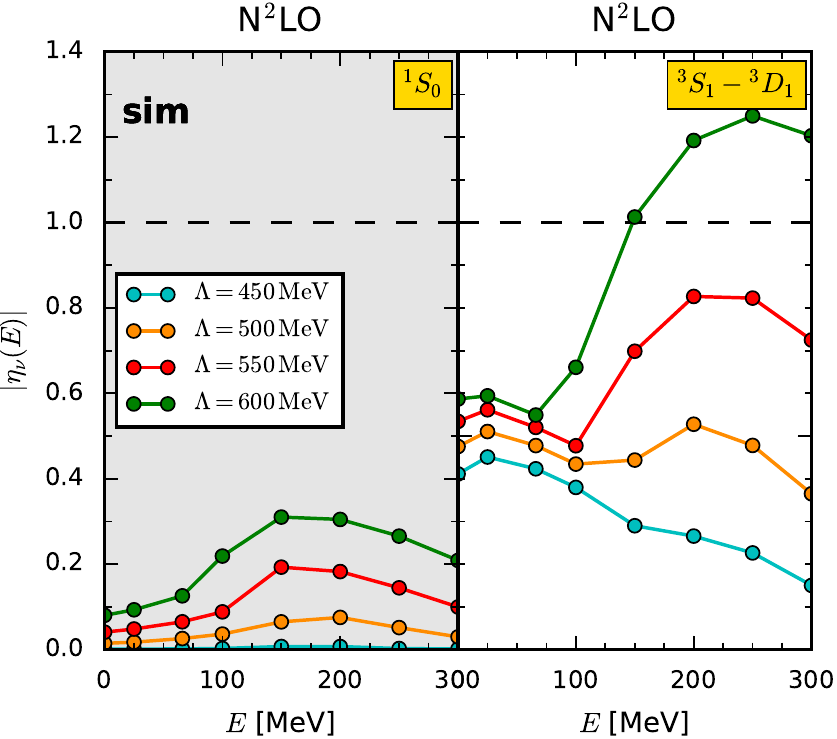}
\caption{\label{fig:rep_WE_sim_290}(Color online)
Magnitude of the repulsive Weinberg eigenvalues for the N$^2$LO sim
potential with $T_\text{rel}=290$~MeV and the cutoff range
$\Lambda=450-600$~MeV as a function of energy $E=0, 25, 66, 100, 150,
200, 250,$ and $300$~MeV in the \oneSzero (left panel) and \threeSDone
channels (right panel).}
\end{figure}

We also examined the $S$-wave repulsive eigenvalues for selected
nonlocal N$^2$LO sim potentials, which are shown in
Fig.~\ref{fig:rep_WE_sim_290}. They are similar to the EKM and EMN
results in the \oneSzero channel, while in the \threeSDone channel the
eigenvalues show a spread in $\Lambda$ as for the N$^2$LO EMN
potential.  In addition, the pattern of energy dependence is different
except for the softest cutoff.

\begin{figure*}[p!]
\includegraphics[page=1,scale=0.97,clip]{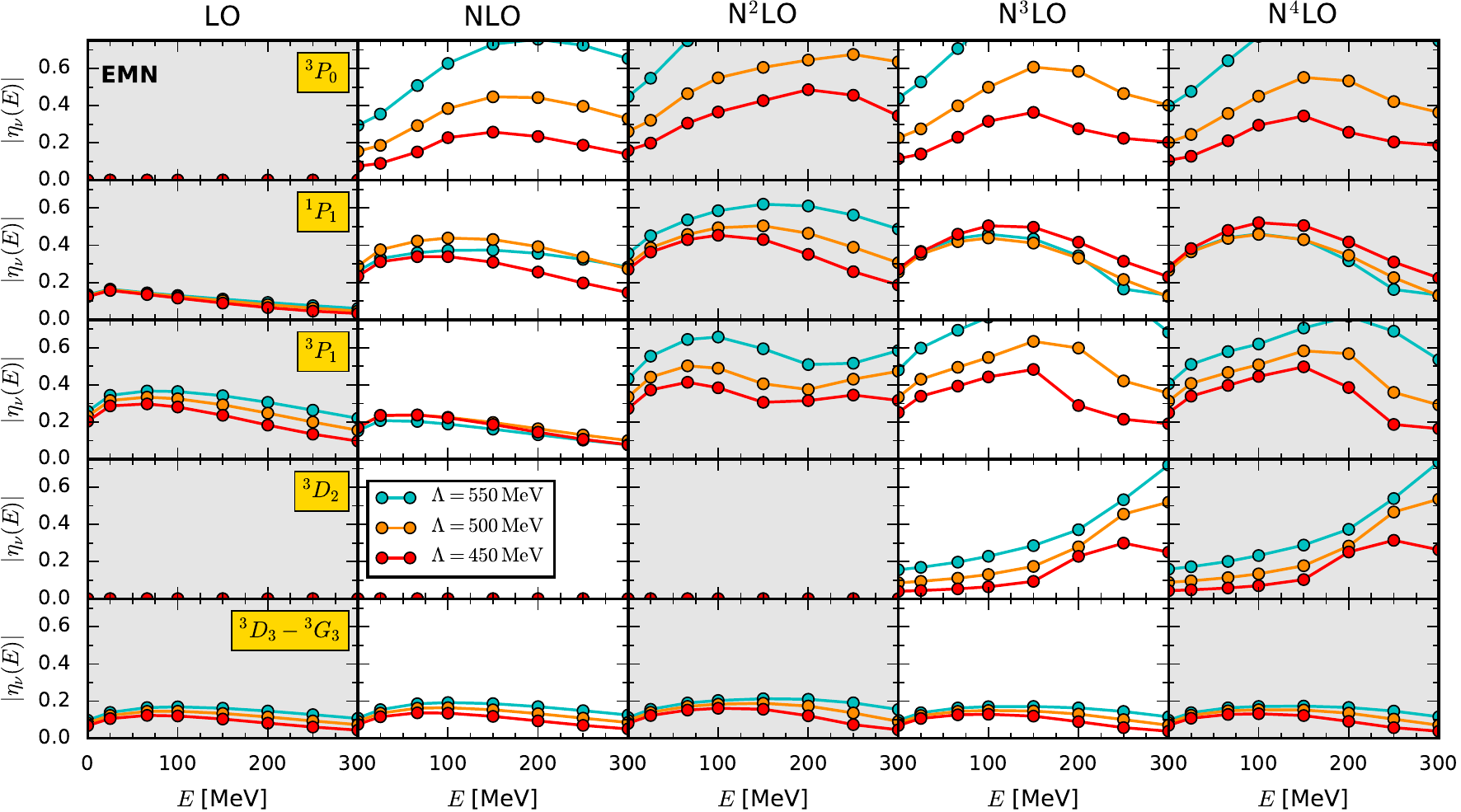}
\caption{\label{fig:rep_WE_EMnew_high_pw}(Color online)
Magnitude of the repulsive Weinberg eigenvalues for the EMN potential up to
N$^4$LO as a function of energy $E=0, 25, 66, 100, 150, 200, 250,$ and $300$~MeV in
different higher partial waves. We show results for momentum-space cutoffs
$\Lambda=450-550$~MeV. Notice that some eigenvalues are partially above the
scale, as we apply the same plot range at all chiral orders and partial waves for better comparison.}
\bigskip
\includegraphics[page=1,scale=0.97,clip]{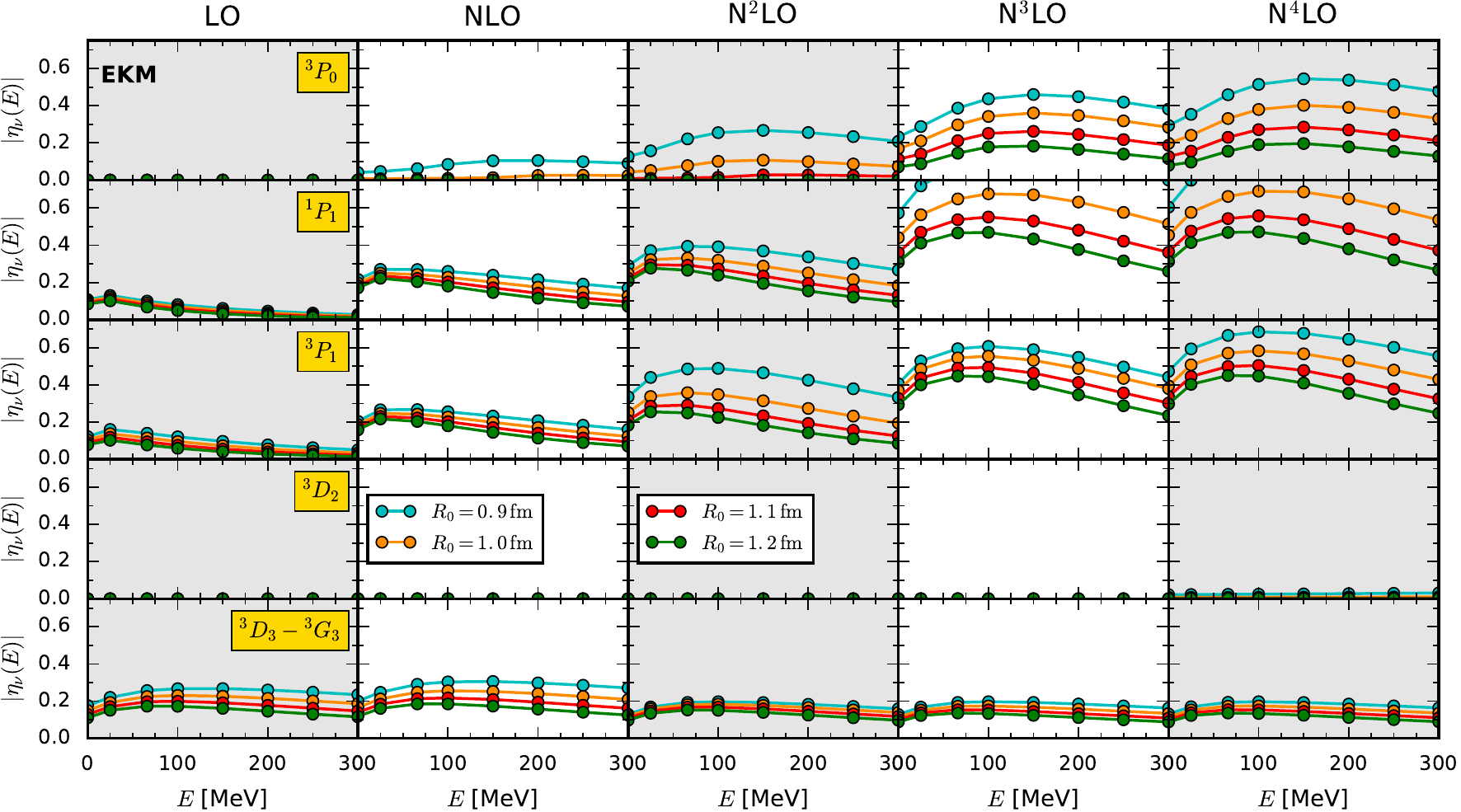}
\caption{\label{fig:rep_WE_semilocal_high_pw}(Color online) Magnitude of the
repulsive Weinberg eigenvalues for the  EKM potential up to N$^4$LO as a
function of energy $E=0, 25, 66, 100, 150, 200, 250,$ and $300$~MeV in different
higher partial waves. We show results for coordinate-space cutoffs
$R_0=0.9-1.2$~fm.}
\end{figure*}

Examples of repulsive eigenvalues in the higher partial waves for the
EMN and EKM potentials are shown in
Figs.~\ref{fig:rep_WE_EMnew_high_pw} and
\ref{fig:rep_WE_semilocal_high_pw}, respectively. In most channels
there are not significant differences. The increases going from
N$^2$LO to N$^3$LO noted for the $S$ waves are present for the EKM
$P$ waves but without the dramatic jumps.  These are only seen for the
EMN potential in the $^3$D$_2$ channel.  The energy dependence of the
repulsive eigenvalues is generally similar even for different
regulators.  However, as noted, the N$^2$LO sim potential shows quite
different energy dependence in the \threeSDone channel as the cutoff
increases.

\begin{figure*}[p!]
\includegraphics[page=1,scale=0.97,clip]{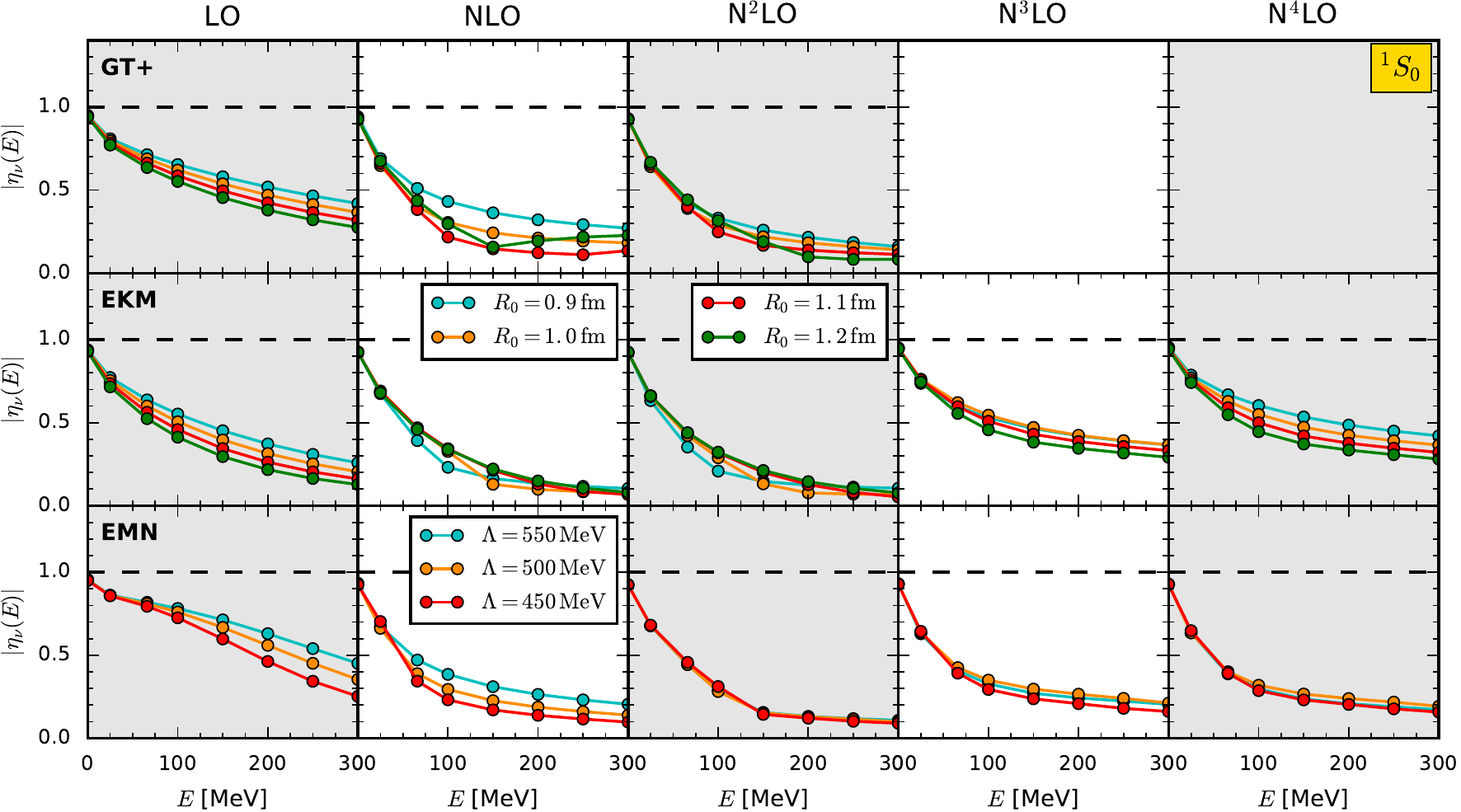}
\caption{\label{fig:att_WE_loc_semiloc_EM_new_1s0}(Color online)
Magnitude of the attractive Weinberg eigenvalues for the GT+  (first row), EKM
(middle row), and EMN potentials (bottom row), as a function of energy $E=0, 25,
66, 100, 150, 200, 250,$ and $300$~MeV in the \oneSzero channel up to the highest chiral
order available, respectively. We show results for coordinate-space cutoffs
$R_0=0.9-1.2$~fm for the GT+ and EKM potential, as well as for momentum-space
cutoffs $\Lambda=450-550$~MeV for the EMN potential.}
\bigskip
\includegraphics[page=1,scale=0.97,clip]{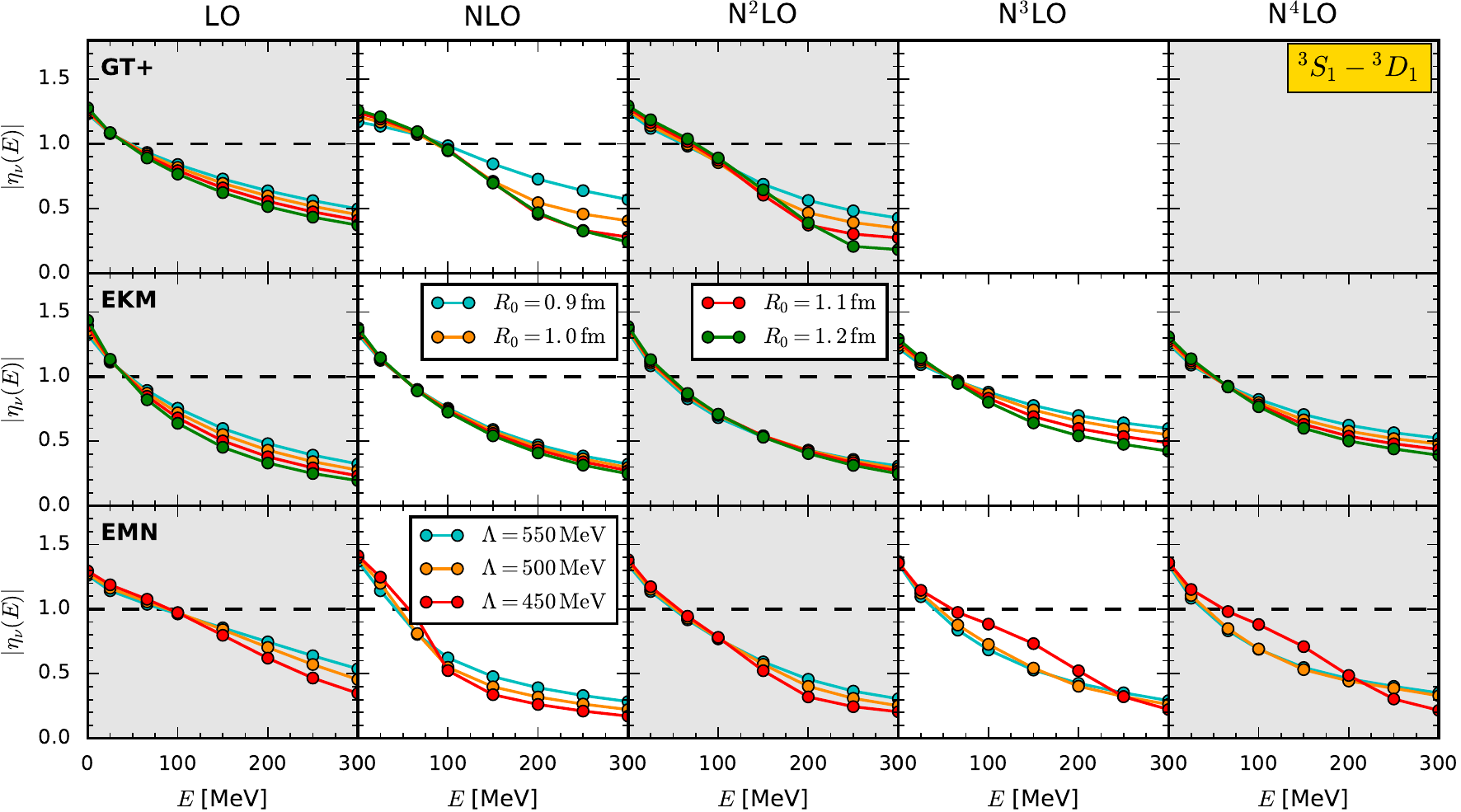}
\caption{\label{fig:att_WE_loc_semiloc_EM_new_3sd1}(Color online) 
Same as Fig.~\ref{fig:att_WE_loc_semiloc_EM_new_1s0} but for the \threeSDone
channel.}
\end{figure*}

The attractive eigenvalues in the \oneSzero and \threeSDone channel
are shown in Figs.~\ref{fig:att_WE_loc_semiloc_EM_new_1s0} and
\ref{fig:att_WE_loc_semiloc_EM_new_3sd1}, respectively, for the GT+,
EKM, and EMN potentials. We find only minor dependence on the cutoff
and nearly the same eigenvalues for all potentials at all chiral
orders. This behavior follows because the magnitude of the attractive
eigenvalues is determined by the shallow or nearly bound state to be
close to 1 at low energies. The energy dependence for all potentials
at all orders and in both channels shows the same fall-off toward
perturbative values.

\begin{figure}[t!]
\includegraphics[]{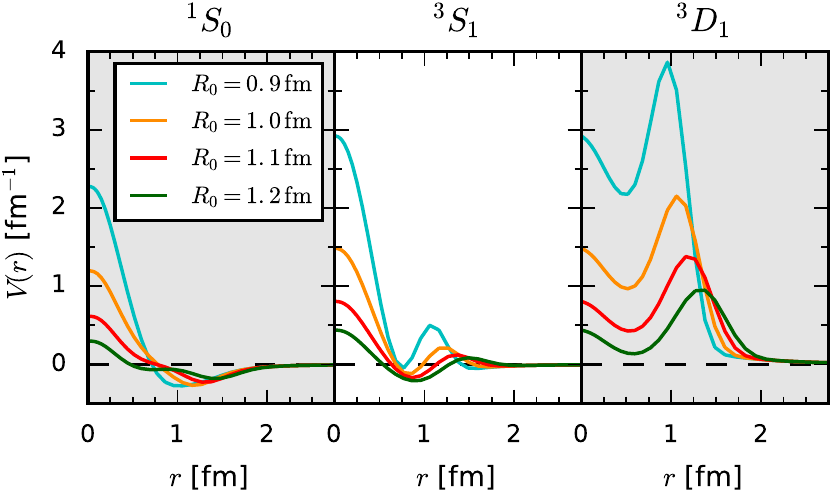}
\caption{\label{fig:Vr_N2LO}(Color online)
Coordinate-space representation of the GT+ potential at N$^2$LO for the 
cutoff range $R_0=0.9-1.2$~fm in the \oneSzero (left), \threeSone
(middle), and \threeDone (right) channels.}
\end{figure}

These many observations illustrate how Weinberg eigenvalues may point
to subtle issues, e.g., with the fitting procedure, but following up
in detail is beyond the scope of this paper.  Instead we give examples
of more general conclusions from consideration of the eigenvalue
systematics:
\begin{itemize}

  \item For the EKM potential, we traced the increased eigenvalues at
    N$^3$LO and N$^4$LO to the new contacts at N$^3$LO.  We observe
    eigenvalues equal to zero for the potential without N$^3$LO
    contacts in the \oneSzero channel, and significantly reduced
    eigenvalues (below 1) in the \threeSDone channel. We conclude
    that the main contribution to the change in magnitude is from the
    contacts at this order.

  \item The repulsion needed to obtain correct phase shifts at high
    energies is provided by contact terms, but how this is realized
    differs between local and nonlocal implementations.  For local
    potentials, the repulsive part is largely built up through the
    energy-independent LECs, because the $q^2$-dependent contacts at
    NLO and beyond are suppressed by at least a factor $r^2$ in
    coordinate space.  This LEC contributes equally at lower energies,
    leading to enhanced eigenvalues at NLO and beyond.  The buildup
    of the short-range repulsion is visible in Fig.~\ref{fig:Vr_N2LO}
    for the N$^2$LO GT+ potential in coordinate space.  In contrast,
    contact terms for the semilocal and nonlocal potentials at NLO and
    beyond also depend on $k^2$, which allows for momentum
    dependence, with large (small) repulsion for higher (lower)
    energies. Here, $\vec{k}=(\vec{p}+\vec{p}')/2$ is the momentum transfer in the exchange channel.

  \item We observed reduced eigenvalues when going from NLO to
    N$^2$LO. This could be due to the improved description of the
    midrange part of the potential as a result of the subleading two-pion
    exchange, entering at N$^2$LO, which requires less fitting into
    the contact parameters at this order.

  \item While one might have guessed that the enhanced repulsive
    Weinberg eigenvalues are due to the low- to high-momentum coupling
    of local regulators, this is actually not the case. This has been
    verified by adding an additional sharp cutoff of
    $\Lambda=4$--$5$~fm$^{-1}$, which leaves the eigenvalues nearly
    unchanged, showing that they are determined by the contributions
    below this cutoff.

\end{itemize}
In general, even when comparing regulators for different potentials
can be quite cumbersome, the Weinberg eigenvalue analysis as a
diagnostic tool offers the possibility to study the perturbativeness,
indicate scheme dependence and possible issues in the fitting
procedure, as well as draw conclusions on the regulator impact.

\begin{figure}[t!]
\includegraphics[page=1,scale=1.0,clip]{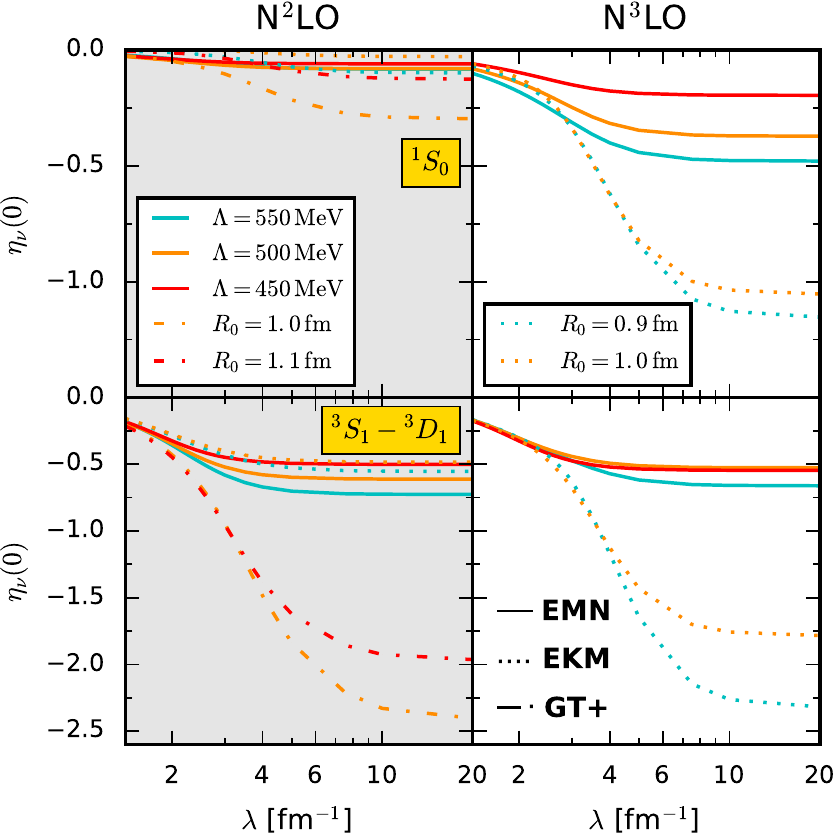}
\caption{\label{fig:WE_SRG_universality}(Color online)
Repulsive Weinberg eigenvalues for the N$^2$LO EMN (solid lines), EKM (dotted lines), and GT+ (dash-dotted lines) and N$^3$LO  EMN (solid lines) and EKM (dotted lines) NN potentials 
 at $E=0$ as a
function of the SRG resolution scale $\lambda$ in the \oneSzero (upper
panels) and the \threeSDone channels (lower panels). For small
$\lambda$, the eigenvalues are in good agreement and exhibit the
universality for potentials evolved to low resolution scales.}
\end{figure}

For a given family of potentials, defined with the same regularization
scheme and constructed with the same fitting protocol, the repulsive
Weinberg eigenvalues reflect the softening of the interaction with
progressively smaller (larger) regulator parameters in momentum
(coordinate) space. This softening can also be realized through an RG
evolution, e.g., via the similarity RG (SRG). In
Fig.~\ref{fig:WE_SRG_universality} we show the eigenvalues at zero
energy in the \oneSzero and \threeSDone channel at N$^2$LO for the
EKM, EMN, and GT+ potentials, as well as at N$^3$LO for the EKM and EMN
potentials as a function of the SRG parameter $\lambda$. The eigenvalues
at large $\lambda$, which correspond to the unevolved (initial)
potentials, exhibit the dramatic jump in hardness from \oneSzero to
\threeSDone for GT+, and in both channels from N$^2$LO to N$^3$LO for
EKM. The jump is much smaller for EMN \oneSzero and no change or even
a softening is observed for EMN \threeSDone.  With evolution to
smaller $\lambda$, all potentials are monotonically softened, with
even the EKM N$^3$LO and GT+ N$^2$LO \threeSDone eigenvalues becoming
perturbative for $\lambda < 4\fmi$, and $\lambda < 3.5 \fmi$,
respectively.

\begin{figure}[t!]
\includegraphics[page=1,scale=1.0,clip]{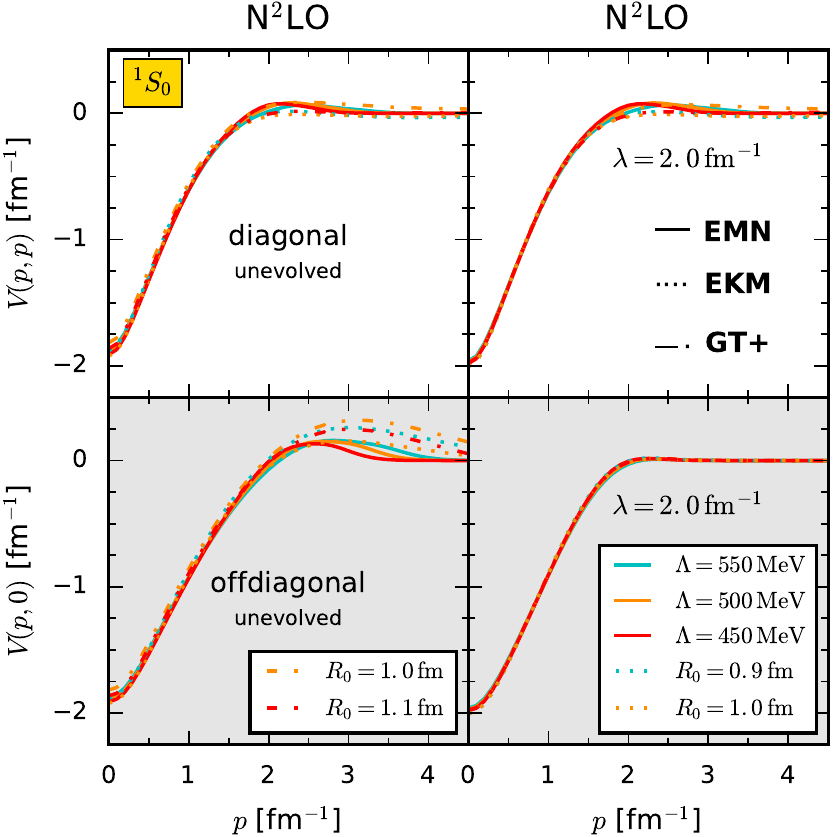}
\caption{\label{fig:pot_SRG_N2LO}(Color online) 
Diagonal (upper row) and offdiagonal (lower row) matrix elements $V(p,p'=p)$, $V(p,p'=0)$, respectively, of the unevolved (left column) and SRG evolved to
$\lambda=2.0$~fm$^{-1}$ (right column) N$^2$LO potentials of EMN (solid
lines), EKM (dotted lines), and GT+ (dash-dotted lines) in momentum space in the \oneSzero
channel. For small $\lambda$ the diagonal elements of all potentials
are again in good agreement.}
\end{figure}

\begin{figure}[t!]
\includegraphics[page=1,scale=1.0,clip]{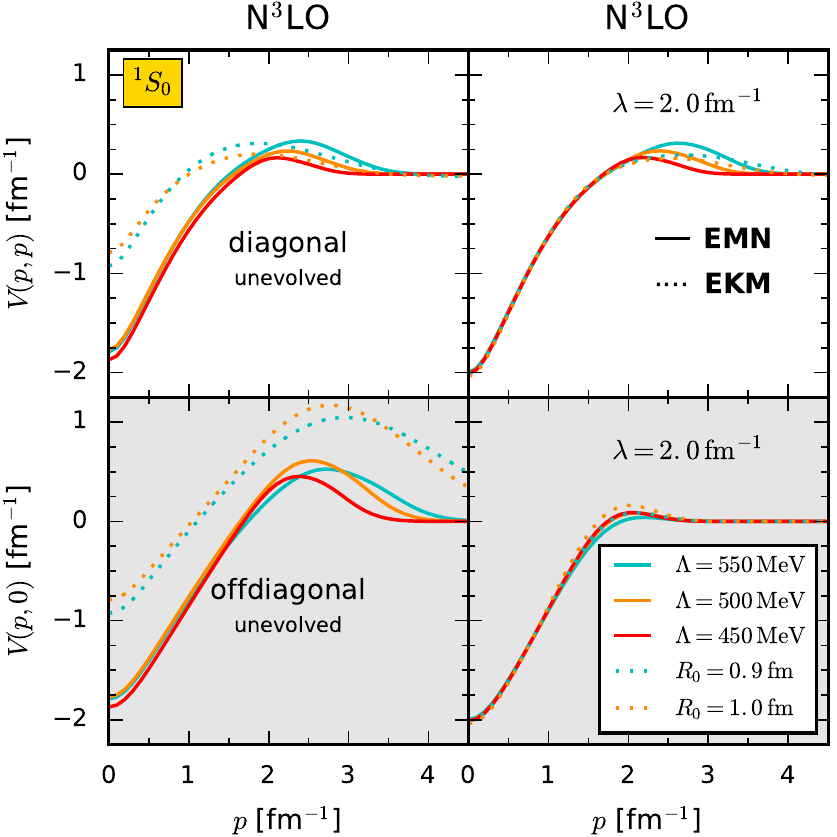}
\caption{\label{fig:pot_SRG_N3LO}(Color online)
Same as Fig.~\ref{fig:pot_SRG_N2LO} but for the EMN and EKM potentials at N$^3$LO.}
\end{figure}

The fine details of the eigenvalue flow mirror the flow of the
potential matrix elements. In Figs.~\ref{fig:pot_SRG_N2LO} and
\ref{fig:pot_SRG_N3LO} we show the unevolved and SRG-evolved diagonal
and off-diagonal matrix elements in the \oneSzero channel for the EMN,
EKM, and GT+ potentials at N$^2$LO, as well as the EMN and EKM potentials at N$^3$LO,
respectively, as functions of the momentum. At N$^2$LO, the relatively
small degree of softening reflects the suppression of off-diagonal
matrix elements, and all matrix elements are quantitatively close for
$\lambda = 2 \fmi$. At N$^3$LO, both diagonal and off-diagonal matrix
elements exhibit a flow toward universal potentials for momenta below $\lambda$.

\section{Summary and outlook}
\label{sec:summary}

In this paper we performed a comprehensive Weinberg eigenvalue
analysis of a representative set of modern NN interactions derived
within chiral EFT. Our results provide insights into the
perturbativeness and scheme dependencies of these interactions.

We find that the attractive eigenvalues, determined by the shallow or
nearly bound states in the \oneSzero and \threeSDone channels, show a
universal behavior for all investigated potentials at all orders in
the chiral expansion. In contrast, the repulsive eigenvalues depend on
specific details such as the regularization scheme, in particular for
the short-range parts of the interaction. This means that the
eigenvalues at different orders in the chiral expansion for a given
class of interactions can behave quite differently. While the GT+
potentials develop large repulsive eigenvalues from LO to NLO, the EKM
potentials remain perturbative up to N$^2$LO and become
nonperturbative only at N$^3$LO and N$^4$LO. We can trace back this
sudden increase at N$^3$LO to the presence of new short-range
couplings at this order. In comparison, the investigated nonlocal
potentials EMN and sim tend to remain more perturbative at all orders.

Moreover, we found that a direct comparison of coordinate-space cutoff
values for the GT+ and EKM interactions can be quite misleading due
to different functional forms of the employed regulators. For example,
we find that a cutoff of $R_0^\text{GT+} = 1.2$~fm essentially
corresponds to $R_0^\text{EKM} \approx 0.8$~fm.  This highlights that
direct comparisons of regulator parameters are not warranted;
alternative ways to compare are given in
Sec.~\ref{sec:NN_interactions}. Finally, we examined the flow to
universality of Weinberg eigenvalues and interaction matrix elements
for the GT+, EKM, and EMN potentials under SRG evolution.

In future work, our analysis can be directly extended to study
regulator artifacts at finite density via in-medium eigenvalues and to
include 3N interactions to assess their impact on perturbativeness.
Furthermore, a comparison of potentials containing delta resonances to
delta-less potentials, which are expected to have different
order-by-order convergence patterns, would be illuminating.  The
applications shown in this paper, including the relation to
phase shifts, suggest that Weinberg eigenvalues can serve as a useful
feedback in fitting potentials by pointing to subtle issues in the
fitting procedure and offering a tool to assess alternative regulator
choices.

\begin{acknowledgments}

We thank J.~E.~Lynn, R.~Machleidt, I.~Tews, and S.~Wesolowski for useful
discussions, and R.~Machleidt also for providing us with the EMN
potentials. C.D. thanks the OSU theory group for the warm hospitality.
This work was supported in part by the European Research Council Grant
No.~307986 STRONGINT, the Deutsche Forschungsgemeinschaft through
Grant SFB~1245, the U.S.\ National Science Foundation under Grant
No.~PHY--1614460, and the NUCLEI SciDAC Collaboration under
U.S.\ Department of Energy Grant \mbox{DE-SC0008533}.  Computational
resources have been provided by the Lichtenberg high performance
computer of the TU Darmstadt.

\end{acknowledgments}

\newpage

\bibliographystyle{apsrev4-1}
\bibliography{strongint}

\end{document}